\documentclass[runningheads]{llncs}

\usepackage{graphicx}
\usepackage{moreverb}
\usepackage[super]{nth}
\usepackage[authoryear,longnamesfirst]{natbib}

\begin{document}

\title{Oxford-style Debates in Telecommunication and Computer Science Education}
\titlerunning{Oxford-style Debates in IT Education}

\author{Marcin Niemiec}
\institute{AGH University of Science and Technology, Mickiewicza 30, 30-059 Krakow, Poland \\
 https://orcid.org/0000-0002-3909-9592 \\
\email{niemiec@agh.edu.pl}}

\maketitle  

\begin{abstract}
Oxford-style debating is a well-known tool in social sciences. Such formal discussions on particular topics are widely used by historians and sociologists. However, when we try to go beyond standard thinking, it turns out that Oxford-style debating can be a great educational tool in telecommunication and computer science. This article presents this unusual method of education at technical universities and in the IT industry, and describes its features and challenges. Best practices and examples of debating are provided, taking into account emerging topics in telecommunications and computer science, such as cybersecurity. The article also contains feedback from IT engineers who participated in Oxford-style debates. All this aims to encourage this form of education in telecommunication and computer science.

\end{abstract}

\section{Introduction}

Oxford-style debating for telecommunication and computer engineers may sound like a strange idea. Indeed, such debates are generally seen as a tool in social sciences – history, economics, political science and sociology. At times they are also be used in natural sciences, especially as individual disciplines intertwine, or in selected interdisciplinary challenges. However, experience shows that this tool may also be useful in engineering education, including telecommunications and computer science. 

Debating has a long history dating back to ancient Greece and Rome. However, the contemporary form of debating has been shaped primarily in England and Scotland. The best-known debating societies active today are the Oxford Union at the University of Oxford
and the Cambridge Union at the University of Cambridge.
Both societies were founded in the first half of the 19th century and many well-known politicians and thinkers have passed through them since then. Debating clubs usually have their own customs and rules for conducting debates. However, some basics are common. This article introduces the specific case of debating: Oxford-style debates.

The rest of the article proceeds as follows. Related work is reviewed in Section 2. In Section 3, the rules of an Oxford-style debate is discussed. Then the advantages of this tool are presented to encourage this form of education. Section 5 explains how to define the proper motion. This section contains also the example motions related to cybersecurity. Feedback form students is presented in Section 6. Section 7 contains advices how to start using Oxford-style debates in telecommunication and computer science education. Section 8 concludes the article.

\section{Related work}

Oxford-style debates in social science is a well-known didactic tool~\citep{sony}. \citet{apply} confirmed that the debate is an important element of teaching, often used as the kick-off to the classroom discussion. Additionally, they provided the result of survey which was designed to determine the effectiveness of the Oxford-style debate as a pedagogical method.

The Oxford-style debates are also used in natural science education~\citep{dent,neuro}. \citet{great} confirmed in the paper ‘\textit{It's no debate, debates are great}’ that debates can be used successfully as an educational tool in different health disciplines. Microbiology~\citep{mbiol}, pharmacy~\citep{phar}, and health care~\citep{health} are the motivating examples. \citet{nurse} showed  the  positive  views  of  nursing  students  towards  classroom  debates.

Oxford-style debates in engineering education are still unusual. Nowadays, engineers rather support participants of debates using technology solutions than take advantage of Oxford-style debates in education process~\citep{Dhillon2016}. However, the first proposals to use this tool for engineers teaching appeared~\citep{7943106}. \citet{7943105} applied the debate into Innoenergy Master school program: \textit{Clean Fossil Fuel and Alternative Energy}. The authors claimed that Oxford-style debate gives  chance  for  students to summarize and recapitulate of the acquired knowledge. Unfortunately, in telecommunications and computer science education the Oxford-style debate is almost completely unknown. In 2021 the author of this paper proposed Oxford-style debates for IT students~\citep{niemiec} during \textit{Lightning Talks} session of the 52nd ACM Technical Symposium on Computer Science Education.

\section{Oxford-style debates}

Oxford-style is not the only type of debating. It is not even the best known, since public and parliamentary debating are more common. However, this kind of debate is an excellent education tool given its well-organized structure with strictly defined rules, including the order and duration of speaking.

\begin{figure}
	\centering
		\includegraphics[scale=.55]{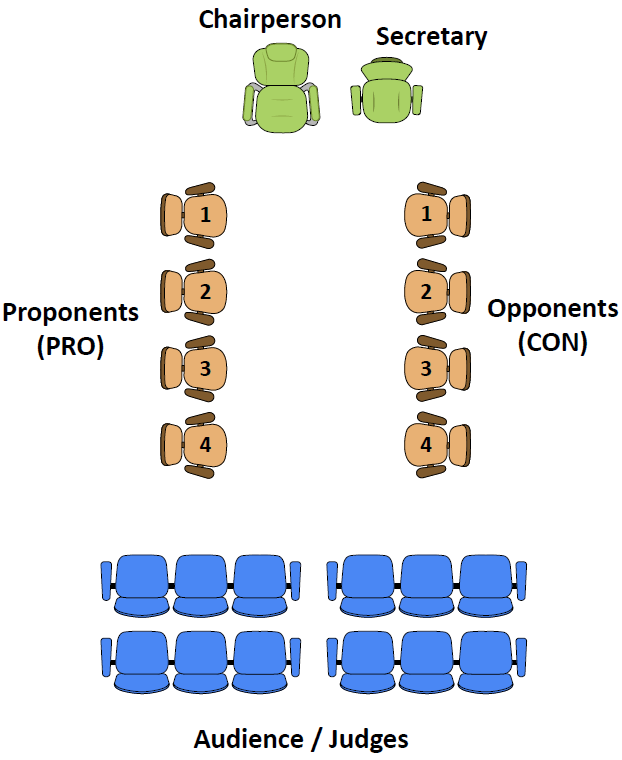}
	\caption{Oxford-style debate scheme.}
	\label{FIG:1}
\end{figure}

Oxford-style debating can be defined as a formal communication process where participants argue for and against a predetermined topic, known as a motion.
The discussion has a competitive format where the participants are divided into two teams: proponents (PRO), who agree with the motion of debate, and opponents (CON), who argue against the motion.
Usually, each group consists of four participants and each individual is assigned a specific role. Proponents and opponents should sit opposite each other, on the model of the British Parliament. The typical layout of the room with participants during Oxford-style debates is shown in Figure \ref{FIG:1}.

The debate is led by a chairperson, more formally known as the Speaker of the House. This individual should be respected by the participants of the debate. The chairperson supervises the course of the discussion, announces speakers, gives the floor and also has the right to take the floor themselves – for example, if the time limit is exceeded or in case of a serious violation of discussion standards. The chairperson may be assisted by the secretary, who is responsible for informing the speakers of the time remaining until the end of their speech (e.g. by tapping lightly on the table a minute before the end of the allocated time). Both the chairperson and the secretary must be impartial and objective throughout the debate.

After introducing the teams and reading the motion, the chairperson gives the floor to the first speaker of the PRO team. Speakers from both sides take the floor alternately, therefore the next speaker is the first representative of the CON team, then the second speaker of the PRO team, etc. The debate ends following the speech by the last speaker from the CON team. Each speaker has a strict limit of time -- usually five minutes -- to convince the audience in favor or against the motion. During the debate, members of the opposing team may ask the speaker questions or give concise information regarding the issue at hand. However, these short interludes must not dominate the speech, thus they usually are limited, e.g. to two per speech. At the end, the audience or chosen representatives (judges) assess and give the result of the debate. Constructive feedback is also given to individual participants, especially if the debate has an educational purpose.

Oxford-style debates are of a competitive nature between two teams, where each speaker of a given team has a specific role (the roles are the same in the PRO and CON teams). 
\begin{itemize}
    \item The first speaker starts the debate by presenting the motion and the team’s standpoint, and by explaining how the team understands this motion. It is often necessary to clarify some keywords. Then, the first speaker outlines the team’s pivot argument to define the framework of the debate.
    \item The role of the second speaker is to develop the factual arguments for or against the motion (supporting the PRO or CON team, respectively). As such, this speaker is responsible for presenting the team's arguments, explaining them in detail and providing evidence.
    \item The third speaker’s presentation should concentrate on the counter-argument. They should try to refute all of their opposition’s arguments, mainly by showing logical errors or wrong examples. Depending on the available time, the third speaker can also complement their own team’s arguments. It is worth mentioning that a new argument may be an accurate counter-argument against the opposing team’s standpoint. 
    \item The last speaker summarizes the debate, focusing on the team’s standpoint. They should recap their team’s understanding of the motion and state the most important arguments, concluding why the speaker’s team should win the debate. 
\end{itemize}

One of the most important points is that the debate is dynamic and speakers should respond to what is currently happening~\citep{conv}. Therefore, sometimes they need to take over the role of another team member, e.g. by completing and validating arguments presented by the previous speaker or by indicating any obvious logical errors in the speech of the opposite team’s speaker. However, speakers must remember that taking on the role of another speaker should not mean they do not fulfill their own role. The basic roles of speakers during an Oxford-style debate are summarized in Table \ref{TB-1}.

\begin{table} 
\caption{The basic roles of speakers during an Oxford-style debate.}\label{TB-1}
\renewcommand{\arraystretch}{1.8} 
\begin{tabular}{p{0.2\linewidth}|p{0.8\linewidth}|}
 \cline{2-2}
 &  \large{\textbf{Role of speaker}} \\  
 \hline
 \multicolumn{1}{ |c| }{ \textbf{\nth{1} speaker}} & Explains the motion and the team’s standpoint. Defines keywords. Outlines the team’s pivot argument. \\
  \hline
  \multicolumn{1}{ |c| }{ \textbf{\nth{2} speaker}} & Presents, explains and provides evidence for their arguments. \\
  \hline
  \multicolumn{1}{ |c| }{ \textbf{\nth{3} speaker}} & Presents counter-arguments and indicates weaknesses of the opposing team’s arguments. \\
  \hline
  \multicolumn{1}{ |c| }{ \textbf{\nth{4} speaker}} & Summarizes the debate and emphasizes key arguments. \\
  \hline
\end{tabular}
\end{table}

Although Oxford-style debates have a competitive character, the fundamental rule is ensuring full respect between opposing team. The discussion must be based on arguments -- speakers can attack arguments but cannot attack individuals. Any argumentation ‘ad personam’ is absolutely prohibited. The chairperson must respond to any inappropriate behavior of participants, such as talking during the debate.

\section{Why debate?}

Oxford-style debating is based on factual arguments. Therefore, the participants improve their understanding related to the topic of the debate. It is essential for all speakers to study the subject in depth to find arguments supporting and opposing their position to prepare for their own speech and possible questions during the debate. Such rigorous self-study emphasizes critical thinking about own opinions. Therefore, the participants -- including the audience who are listening the debate – take a broad view of the given subject without ignoring different points of view.

However, learning to find facts and formulate arguments by means of independent research is only the first step of preparations for the discussion. After that, all team members must work together to prepare a coherent line of argumentation. Each speaker is responsible for the whole team and should be ready to fulfill their role. This makes Oxford-style debates an excellent tool for developing efficient teamwork – a crucial skill for modern telecommunication and computer engineers.

Another reason for hosting Oxford-style debates for students is to help them improve their soft skills such as communication, social skills and emotional intelligence~\citep{cometa}. Such skills are important not only for IT engineers but also for future entrepreneurs, who will use soft skills in everyday business practice. It is not without reason that Oxford-style debating is sometimes known as ‘the school of leaders’.

Each speaker during the debate has just a few minutes to convince the audience/judges of their team’s standpoint. Thus, the structure of the speech should be logical and coherent to present arguments in comprehensive way. In addition to verbal communication, speakers also learn the important but difficult (especially for engineers) elements of non-verbal communication such as good diction, voice modulation, clear articulation, convincing body language, gesticulation, eye contact with the audience, etc. Such techniques help engineers improve the effectiveness of communication and enliven public presentations (e.g. during business meetings in the future). Active participation in debates develops negotiation and persuasion skills. Additionally, students learn how to manage and reduce their own stress level during public speeches.

Holding Oxford-style debates also has many benefits for teachers and tutors. They provide them with an opportunity to expand their own knowledge of the subject. Motivated students frequently enrich the raised arguments with relevant examples from their personal experience. Additionally, this innovative teaching method helps teachers and tutors build good relationships with their students.

\section{Debates on computer science topics}

One of the key elements of a successful Oxford-style debate is defining the motion. A typical motion of an Oxford-style debate is an affirmative sentence. Usually, it concerns a certain social or political problem, although it is equally possible to propose interesting motions concerning telecommunications and computer science. Teachers and tutors should follow certain rules when defining such motions. 

Firstly, it is necessary to make sure that the proposed motion does not put one of the teams in a losing position in advance. Therefore, a properly defined motion must not be obvious -- participants in the debate should be able to find a similar number of arguments supporting and opposing it. Secondly, motions concerning certain hot topics in telecommunications or computer science will be motivating for speakers and more interesting for the audience. Thirdly, it is good practice for the motion to be a short, unambiguous statement so that each team understands and interprets it in a similar way. Having a unequivocal and explicit motion reduces the risk that members of each team do not discuss it with each other and each team simply presents their point of view, because presented arguments will lead such a debate in two different directions.

Defining valid and interesting motions is certainly a challenge; however, given sufficient interest and research, there are plenty of topics suitable for a valuable debate related to telecommunications and computer science. 

We will now share our experience in defining motions for Oxford-style debates. Table \ref{TB-2} presents example motions in the cybersecurity area. They were defined for Oxford-style debates held as part of a seminar for master’s degree students in IT. Some of the motions (4, 5 and 7) were successfully discussed during the Winter School of Cybersecurity -- a joint initiative between academia and industry.

\begin{table} 
\caption{Example motions in the cybersecurity area.}\label{TB-2}
\renewcommand{\arraystretch}{1.8} 
\begin{tabular}{p{0.04\linewidth}|p{0.94\linewidth}|}
 \cline{2-2}
 & \large{\textbf{Motion}} \\  
 \hline
 \multicolumn{1}{ |c| }{\textbf{1}} & \textit{Opening the source code increases the security level of the application} \\
  \hline
  \multicolumn{1}{ |c| }{\textbf{2}} &  \textit{COVID-19 pandemic increased security level in cyberspace} \\
  \hline
  \multicolumn{1}{ |c| }{\textbf{3}} & \textit{Law enforcement agencies should be able to read confidential messages sent by citizens}  \\
  \hline
  \multicolumn{1}{ |c| }{\textbf{4}} &  \textit{Former hackers are the best candidates for security specialist positions} \\
  \hline
  \multicolumn{1}{ |c| }{\textbf{5}} &  \textit{Anonymization of Internet users (e.g. using the Tor network) should be prohibited} \\
  \hline
  \multicolumn{1}{ |c| }{\textbf{6}} &  \textit{Research in the IT field which may give rise to global threats 
  should be regulated by law} \\
  \hline
  \multicolumn{1}{ |c| }{\textbf{7}} &  \textit{Universities should not teach students about hacking techniques} \\
  \hline
\end{tabular}
\end{table}

\section{Feedback from students}

The assessment of the proposed didactic tool by participants is crucial. Therefore, from 2020 to 2022 we were asking IT engineers for feedback. Participants of selected seminars (master’s degree students) filled anonymous survey. Each respondent actively participated in Oxford-style debates as a speaker (at least twice) and a judge (at least once). 

It is worth mentioning that approx. 70\% of respondents had heard about Oxford-style debates before. However before the seminar they did not recognize debates as an interesting element of engineering education. After debates the students changed opinion: 80\% of participants agreed on the statement that the debates help to become a better IT engineer. Additionally, all participants were asked to rate the debates as an educational tool in telecommunication and computer science on a scale from $0$ (completely useless) to $10$ (very useful). The overall score of Oxford-style debates as a teaching tool in engineering education is $9$.

\begin{figure}[!tbp]
  \centering
  \begin{minipage}[b]{0.495\textwidth}
    \includegraphics[width=\textwidth]{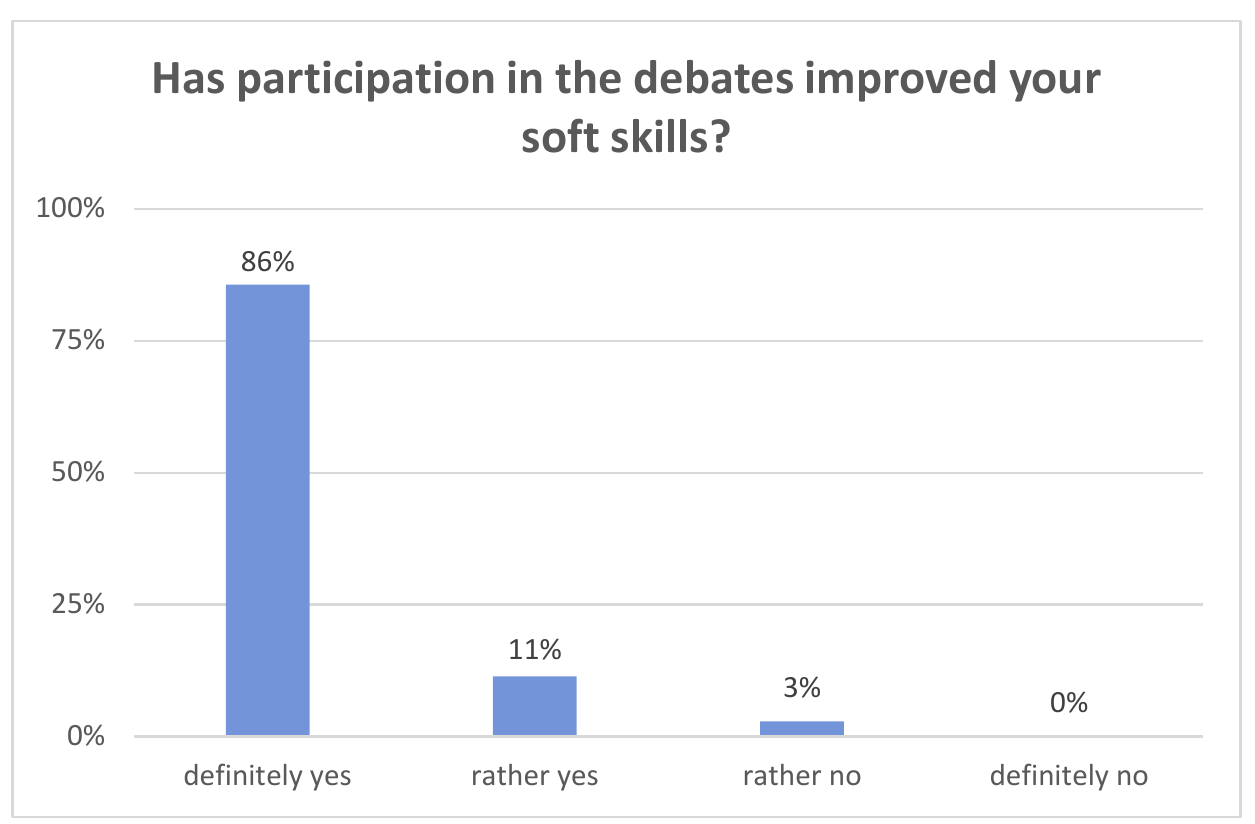}
    \caption{Feedback: soft skills.}
    	\label{FIG:4}
  \end{minipage}
  \hfill
  \begin{minipage}[b]{0.495\textwidth}
    \includegraphics[width=\textwidth]{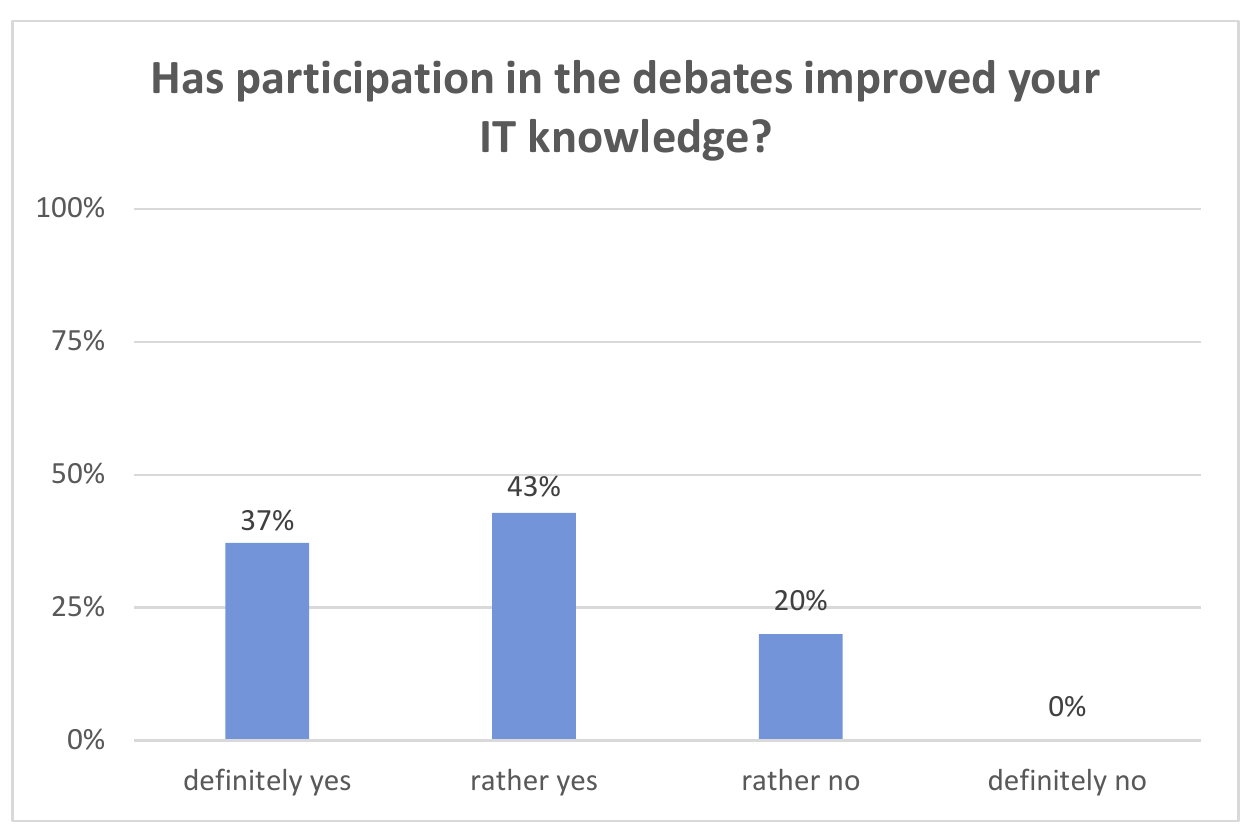}
    \caption{Feedback: IT knowledge.}
    	\label{FIG:5}
  \end{minipage}
\end{figure}

\begin{figure}[!tbp]
  \centering
  \begin{minipage}[b]{0.495\textwidth}
    \includegraphics[width=\textwidth]{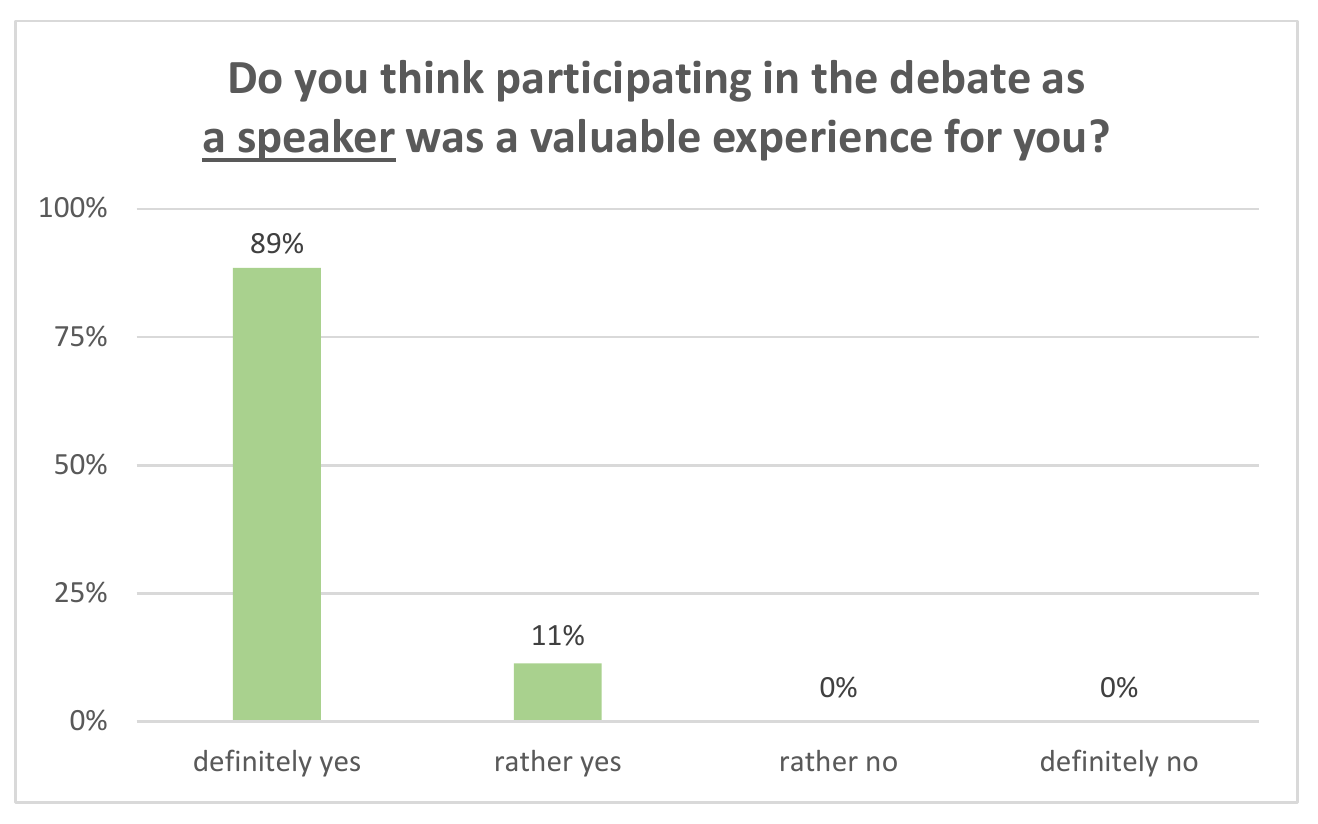}
    \caption{Role in debate: a speaker.}
    	\label{FIG:2}
  \end{minipage}
  \hfill
  \begin{minipage}[b]{0.495\textwidth}
    \includegraphics[width=\textwidth]{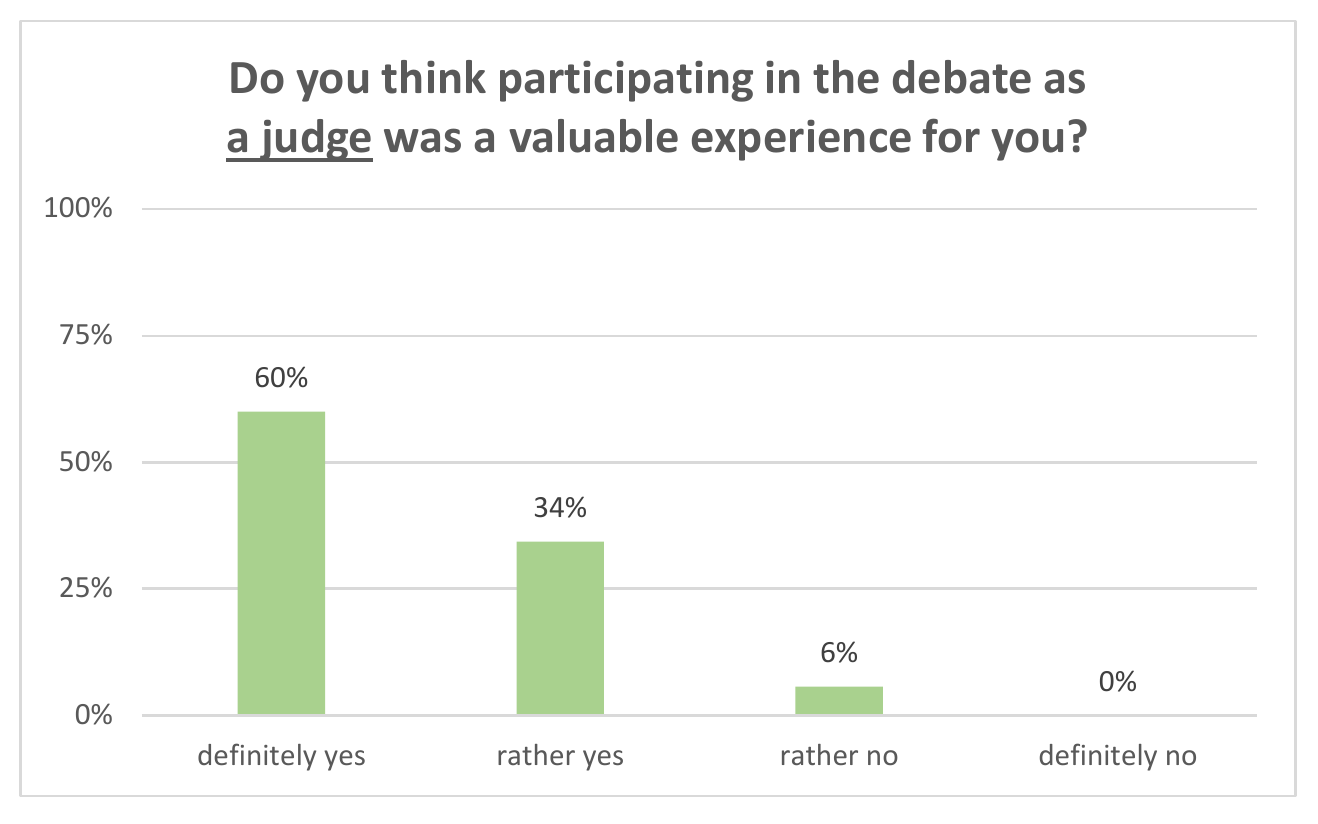}
    \caption{Role in debate: a judge.}
    	\label{FIG:3}
  \end{minipage}
\end{figure}

We also wanted to ask why it is worth to debate. The students answered if they improved soft-skills and IT knowledge thanks to the Qxford-style debates. The very motivating results are presented in Figures  \ref{FIG:4} and \ref{FIG:5}. 
Additionally, the students were asked to assess the roles in debates. The respondents recognized both roles as valuable, however participation as an active speaker is assessed higher than the judge role. The answers are presented in Figures \ref{FIG:2} and \ref{FIG:3}.

\begin{figure}[!tbp]
  \centering
  \begin{minipage}[b]{0.495\textwidth}
    \includegraphics[width=\textwidth]{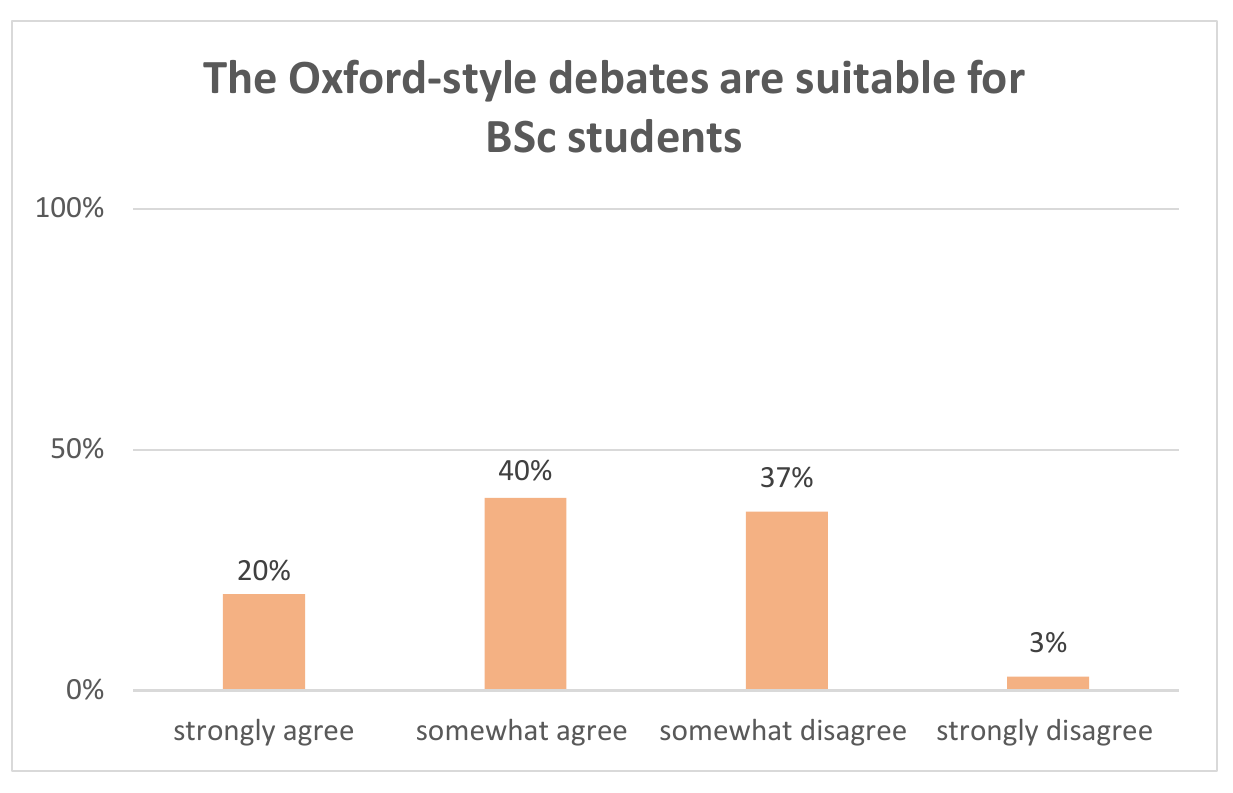}
    \caption{Debates for BSc students.}
    	\label{FIG:6}
  \end{minipage}
  \hfill
  \begin{minipage}[b]{0.495\textwidth}
    \includegraphics[width=\textwidth]{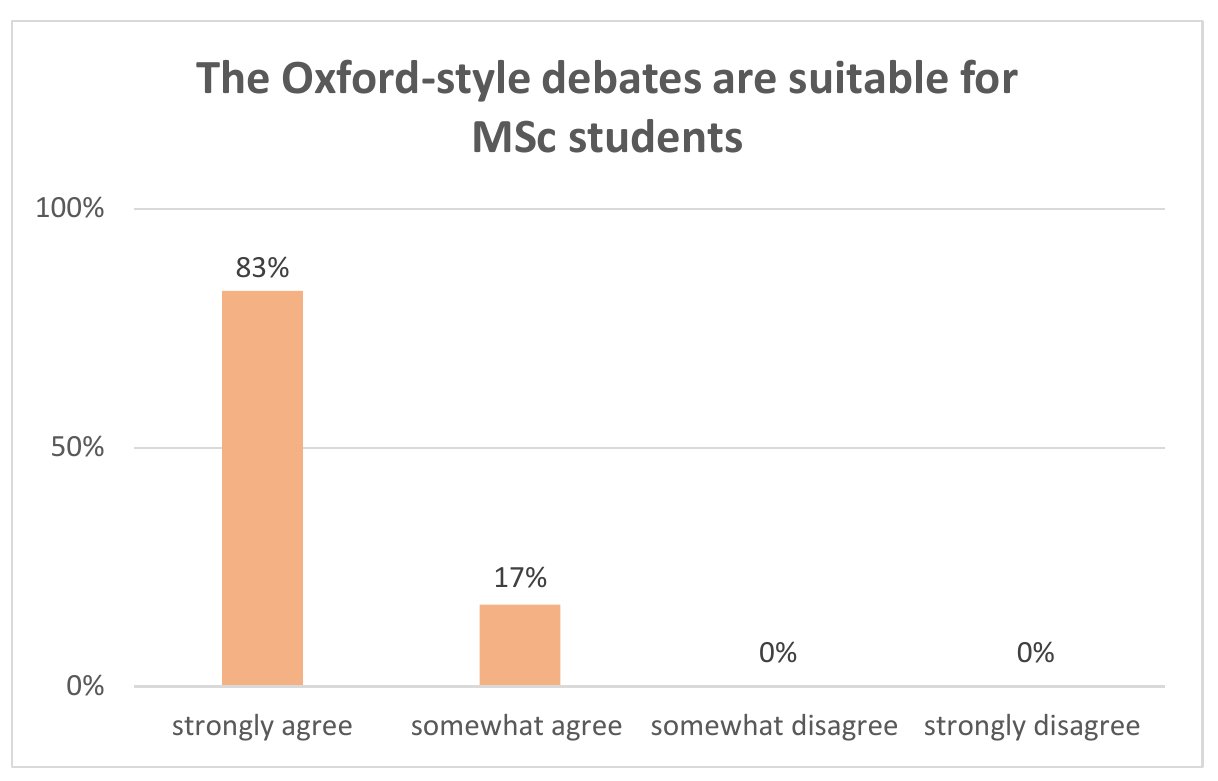}
    \caption{Debates for MSc students.}
    	\label{FIG:7}
  \end{minipage}
\end{figure}

Finally, we asked respondents (master’s degree students) if the Oxford-style debates are suitable for BSc and MSc students. The responses presented in Figures \ref{FIG:6} and \ref{FIG:7} confirm that this didactic tool is more suitable for MSc students than for candidates for engineers.

\section{Getting started}

Oxford-style debates are relevant to BSc, MSc and PhD students. However, debating is particularly recommended for master’s degree students, since they are most likely to benefit from this education tool. Also, students at this stage of education have a strong technical background, which means debating allows them to broaden their horizons and use the knowledge they already have. It is worth mentioning that not all forms of teaching are suitable for Oxford-style debates, with seminars being particularly well suited. Therefore, for beginners, debates are a great didactic tool during master’s degree seminars. They are also an excellent activity for extracurricular events, e.g. summer schools. 

Each debate requires two teams (eight speakers). The rest of students can be the judges, who assess and give the result of the debate. Therefore, Oxford-style debates can be used in various class size. However, different class sizes influence the utility of this didactic tool, because the debate is most beneficial to the active speakers. Therefore, Oxford-style debates are rather recommended to smaller groups of students than larger groups.

When introducing debates to education, it may be advisable to start with simplified rules of discussion. For example, it may be better to reduce the number of questions during the debate or avoid additional elements such as ‘ad vocem’ (a short additional counter-speech following the speaker's speech). Simple debating rules also make the experience easier for beginner speakers. 
An extremely important factor of successful debates is the motion: define an appropriate and interesting motion, and you’re halfway there. Debating interesting motions will feel more like fun than an obligation for students. Of course, the motion should be announced at least a few days before the debate to allow the teams enough time to prepare factual arguments and build a cohesive team.

It is also worth considering to allocate individuals to teams at random. Again, such an arrangement should be announced at least a few days before the debate (i.e. when the motion is announced) to ensure the competition is even and fair. Random selection of participants also helps group integration, because students usually tend to work on projects with the same small teams. Additionally, it provides a scenario typical of business projects, where graduates do not have a choice of who they work with. As a result, this method of selecting team members improves adaptability in new environments and develops their teamworking skills.

In general, it is natural that the teacher/tutor acts as the chairperson. However, it is advisable to watch some example debates before the first supervised discussion; there is plenty of material online to get started. If the chairperson can manage how long each speaker talks for, it may not be necessary to appoint a secretary. It is not essential for an audience to be present during the debate, but it is highly recommended. An audience is an additional motivation for speakers and bolsters competition between the teams. Additionally, the audience can be treated as formal judges of Oxford-style debates, where each participant has their own and equal voice. If the group of students is not large and additional attendees are needed to assess the debate, you can choose invite other colleagues. Finding volunteers is usually not a problem, since people are curious about this unusual form of teaching. 

Once the last speaker finishes, the chairperson calls the vote and announces the result of the debate (based on the assessments of the judges/audience). The chairperson also has a single vote, although it is important that it is not used to decide on the winner. It is worth emphasizing that personalized feedback to each speaker is far more important than the actual result of the competition. It improves the educational value of the debate and helps participants hone their skills. Of course the feedback must be motivating and encouraging further participation in debates. Additionally, it may be useful to have prizes for the winners, even if they are token. Well conducted Oxford-style debates are bound to be enjoyable for students and teachers alike.

\section{Conclusions}

Oxford-style debates are a universal educational tool for students of engineering and social sciences. Although this may come as a surprise to many academic teachers and IT industry experts, in fact this form of debating supports professional education and soft skills training, teamwork and individual study, and verbal and non-verbal communication skills. All of these abilities are very important for IT engineers. The well-organized structure and competitive character makes this educational tool useful in academic study (especially for postgraduate students) as well as at special events (e.g. summer schools). 

It does not take much to start the journey into the world of Oxford-style debates: a short introduction of basic rules to the participants, an interesting motion for discussion and a dose of enthusiasm. Oxford-style debates are highly recommended as a tool in telecommunication and computer science education.

\end{document}